\documentclass[aps,pre,reprint,superscriptaddress]{revtex4-1}

\usepackage{color}
\usepackage{latexsym}
\usepackage{amssymb}
\usepackage{amsmath}  
\usepackage{graphicx}
\usepackage{empheq} %for boxes with multline or align

\usepackage{soul}  %the command is \textst{text} 
\usepackage{empheq} %for boxes with multline or align
\usepackage{tabulary}
\usepackage{tabularx}
\def\rmd{{\mathrm{d}}}
\def\rme{{\mathrm{e}}}
\def\SItext{{\em Supporting Information}}

\newcommand{\eqs}{\;\!}

\newcommand{\A}{_\mathrm{A}}
\newcommand{\B}{_\mathrm{B}}
\newcommand{\BB}{_{\mathrm{B}*}}

\newcommand{\AB}{_\mathrm{A(B)}}
\newcommand{\BA}{_\mathrm{B(A)}}

\newcommand{\Rg}{R_\mathrm{g}}
\newcommand{\kB}{k_\mathrm{B}}
\newcommand{\DG}{\Delta G}
\newcommand{\DGbar}{\Delta \bar{G}}
\newcommand{\Kvol}{K_\mathrm{vol}}

\newcommand{\ktot}{k_\mathrm{tot}}
\newcommand{\kR}{k_\mathrm{R}}
\newcommand{\kDA}{k_{D\A}}
\newcommand{\kDB}{k_{D\B}}

\usepackage{soul}  % for strikethrough font

\begin{document}

%\preprint{}

\title{Catalyzed Bimolecular Reactions in Responsive Nanoreactors}



\author{Rafael Roa}
\email[]{rafael.roa@helmholtz-berlin.de}
\affiliation{Institut f\"ur Weiche Materie und Funktionale Materialien, Helmholtz-Zentrum Berlin f\"ur Materialien und Energie, 14109 Berlin, Germany}

\author{Won Kyu Kim}
\affiliation{Institut f\"ur Weiche Materie und Funktionale Materialien, Helmholtz-Zentrum Berlin f\"ur Materialien und Energie, 14109 Berlin, Germany}

\author{Matej Kandu\v{c}}
\affiliation{Institut f\"ur Weiche Materie und Funktionale Materialien, Helmholtz-Zentrum Berlin f\"ur Materialien und Energie, 14109 Berlin, Germany}

\author{Joachim Dzubiella}
\email[]{joachim.dzubiella@helmholtz-berlin.de}
\affiliation{Institut f\"ur Weiche Materie und Funktionale Materialien, Helmholtz-Zentrum Berlin f\"ur Materialien und Energie, 14109 Berlin, Germany}
\affiliation{Institut f\"ur Physik, Humboldt-Universit\"at zu Berlin, 12489 Berlin, Germany}

\author{Stefano Angioletti-Uberti} 
\email[]{s.angioletti-uberti07@imperial.ac.uk}
\affiliation{Department of Materials, Imperial College London, London SW7 2AZ, UK}
\affiliation{Beijing Advanced Innovation Centre for Soft Matter Science and Engineering, Beijing University of Chemical Technology, 100029 Beijing, PR China}


\date{\today}

\begin{abstract}

We describe a general theory for surface-catalyzed {\it bimolecular} reactions in responsive nanoreactors, catalytically active nanoparticles coated by a stimuli-responsive `gating' shell, whose permeability controls the activity of the process. We address two archetypal scenarios encountered in this system: The first, where two species diffusing from a bulk solution react at the catalyst's surface; the second where only one of the reactants diffuses from the bulk while the other one is produced at the nanoparticle surface, e.g., by light conversion. We find that in both scenarios the total catalytic rate has the same mathematical structure, once diffusion rates are properly redefined. Moreover, the diffusional fluxes of the different reactants are strongly {\it coupled}, providing a richer behavior than that arising in unimolecular reactions. We also show that in stark contrast to bulk reactions, the identification of a limiting reactant is not simply determined by the relative bulk concentrations but controlled by the nanoreactor shell permeability. Finally, we describe an application of our theory by analyzing experimental data on the reaction between hexacyanoferrate (III) and borohydride ions in responsive hydrogel-based core-shell nanoreactors. 

\end{abstract}

%\keywords{}

\maketitle

%%%%%%%%%%%%%%%%%%%%%

\section{Introduction}

Responsive nanoreactors are an emerging and promising new molecular technology for nanocatalysis in which catalyst nanoparticles are confined in hollow nanostructures by permeable shells that can be used to shelter and control the catalytic processes.
In particular,  the catalysis can be made selective and responsive if the shell differentiates among molecules and if the shell permeability can be modulated by external stimuli~\cite{Petrosko:2016he,Stuart:2010hu,Campisi:2016hl,Lu:2011bi,CarregalRomero:2010gp,Herves:2012fp,Wu:2012bx,Jia:2016cy,Prieto:2016bj,Gaitzsch:2015kr,Vriezema:2005fx,Renggli:2011if,Tanner:2011jf,Guan:2011eva}.
These nanoreactors can be used for a large variety of applications, ranging from analytical tools to study chemical reactions~\cite{Petrosko:2016he,Lu:2011bi,CarregalRomero:2010gp,Herves:2012fp,Wu:2012bx,Jia:2016cy,Prieto:2016bj,Vriezema:2005fx,Renggli:2011if,Campisi:2016hl,Stuart:2010hu,Gaitzsch:2015kr} to biosensors for the diagnosis of diseases~\cite{Vriezema:2005fx,Renggli:2011if,Tanner:2011jf,Gaitzsch:2015kr,Guan:2011eva}.
Examples of natural nanoreactors are lipid-based membranes ({e.g.} liposomes), cage-like proteins ({e.g.} ferritins), protein-based bacterial microcompartments, and viruses~\cite{Vriezema:2005fx,Renggli:2011if,Tanner:2011jf,Liu:2016gf}.
Artificial nanoreactors (based on spherical polyelectrolyte brushes, dendrimers, ligands, or even DNA) are simpler than the natural ones and thus easier to control  for targeted applications~\cite{Lu:2011bi,CarregalRomero:2010gp,Herves:2012fp,Wu:2012bx,Jia:2016cy,Prieto:2016bj,Vriezema:2005fx,Renggli:2011if,Montolio:2016jy,Zinchenko:2016jy,Gaitzsch:2015kr}. 

In recent years, nanoreactors containing metal nanoparticles coated with stimuli-responsive polymers have emerged as a promising catalytic system~\cite{Lu:2011bi,Herves:2012fp,Wu:2012bx,Prieto:2016bj,Lu:2006cr,Zhang:2010hx,ContrerasCaceres:2008hk,CarregalRomero:2010gp,Jia:2016cy,Li:2016bs}.
Two are the key roles of the polymer shell. On the one hand, the shell acts as a carrier that protects nanoparticles from aggregation and hinders chemical degradation processes, { e.g.} oxidation \cite{Jia:2016cy}. On the other hand, the polymer ability to switch between states with different physicochemical properties upon changes in environmental parameters, {e.g.}\ temperature, pH, or concentration of certain solutes, provides a handle to actively control the nanoreactor's catalytic properties. \\
A well-studied archetypal {\it active} carrier system is based on poly(N-isopropylacrylamide) (PNIPAM) hydrogels~\cite{Lu:2006cr,Zhang:2010hx,Lu:2011bi,Herves:2012fp,Wu:2012bx,ContrerasCaceres:2008hk,CarregalRomero:2010gp,Jia:2016cy,Li:2016bs}. Here, the shell is in a swollen hydrophilic state at low temperature, but sharply collapses into a hydrophobic state above the critical solution temperature~\cite{Pelton:2000vo}. Examples of catalytic reactions in aqueous solution studied in this system are the reductions of nitrobenzene, 4-nitrophenol, or hexacyanoferrate (III) by borohydride ions~\cite{Lu:2006cr,Herves:2012fp,Wu:2012bx,Zhang:2010hx} and the decomposition of methyl orange under visible light~\cite{Jia:2016cy}.

All the aforementioned examples deal with surface-catalyzed {\it bimolecular} reactions, a very common type. 
In the strictly unimolecular limit (as, {e.g.}, in enzyme kinetics~\cite{Szabo:2012ib}), a single reactant transforms into a product once in the proximity of the nanoparticle surface.
In this latter case, the total catalytic rate (reciprocal of the catalytic time) is calculated by the well-known, exact expression $\ktot^{-1}=k_D^{-1}+\kR^{-1}$, where $k_D$ and $\kR$ are the diffusion and the surface reaction rates, respectively~\cite{Berg:1985ea,Herves:2012fp,Wu:2012bx,AngiolettiUberti:2015go}.
Unimolecular reactions can be diffusion- or surface-controlled if $k_D\ll \kR$ or $k_D\gg \kR$, respectively. If both rates are comparable in magnitude, the reaction is termed diffusion-influenced. Analogously, a reaction is diffusion- or surface-controlled if $\mathrm{Da}_\mathrm{II}\gg1$ or $\mathrm{Da}_\mathrm{II}\ll1$, where $\mathrm{Da}_\mathrm{II}=\ktot/k_D$ is the second Damk\"ohler number~\cite{Herves:2012fp}.

As pointed out before~\cite{CarregalRomero:2010gp,Herves:2012fp,Wu:2012bx, AngiolettiUberti:2015go}, pseudo-unimolecular surface-catalyzed reactions in responsive nanoreactors can be described by combining a thermodynamic two-state model for the polymer volume transition with the appropriate reaction-diffusion equations.  In particular, the important effect of a change in the local permeability on the reactants approach to the catalyst's surface can be described by theory of diffusion through an energy landscape,\cite{Wu:2012bx, AngiolettiUberti:2015go, Galanti:2016bz} in the spirit of Debye--Smoluchowski diffusion-controlled rate theory~\cite{Smoluchowski:1917wh,Debye:1942cv,Wilemski:1973ki,Calef:1983vd,Hanggi:1990en}.
This theoretical framework for pseudo-unimolecular reactions qualitatively rationalizes the large and sharp variations in catalytic rate observed in the relevant experiments~\cite{Herves:2012fp,Wu:2012bx,AngiolettiUberti:2015go,Jia:2016cy}. 
To this end, it was implicitly assumed that bimolecular reactions can be treated as pseudo-unimolecular when one of the reactants is in large excess with respect to the other.
This assumption is correct for reactions in a bulk solution. However, when considering nanoreactors care should be taken since in these systems it is not the bulk concentration that matters, but the reactant concentration at the nanoparticle surface where the reaction can take place. 
The latter is strongly influenced by the shell permeability $\mathcal P$, defined as the product of reactant partitioning and diffusivity~\cite{Palasis:1992ew}, and can thus strongly differ from the bulk value. Due to the responsive nature of the gating shell of nanoreactors, this dependence crucially implies that the identity of the limiting reactant can switch upon a change in the external stimulus.
As we shall see later, failure to recognize this fact can lead to very large discrepancies between the exact and the approximate rate.

%%%%%%%%%%%%%%%%FIG
\begin{figure*}[t!]
\begin{center}
\includegraphics[width=0.87\linewidth]{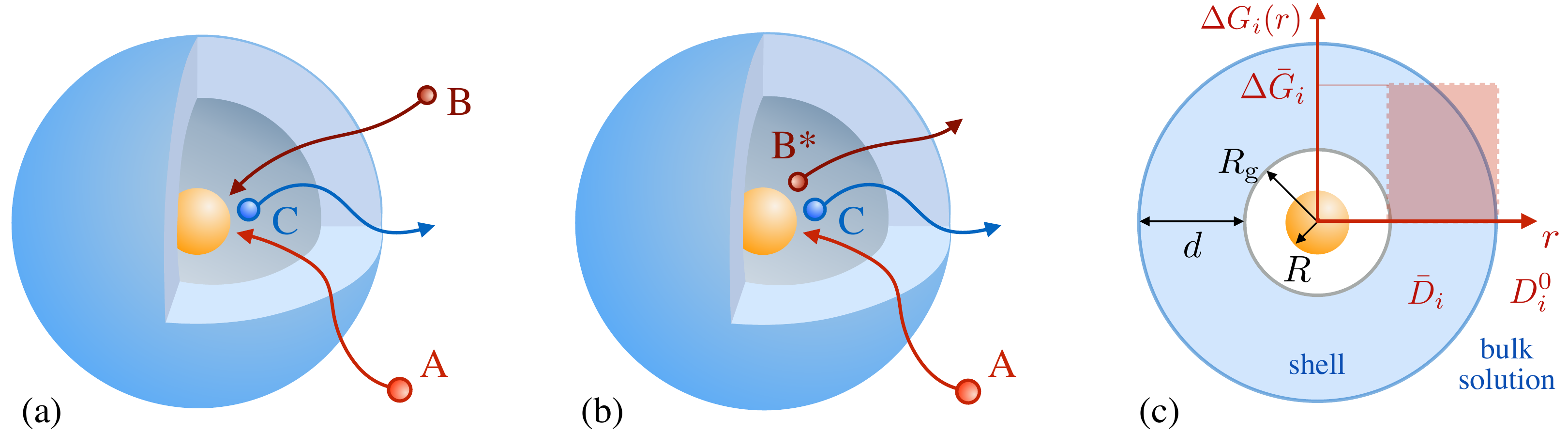}
\caption{Bimolecular reactions in yolk-shell nanoreactors. 
(a) Two reactants, A and B, diffusing from a bulk solution, generate a product, C, in the proximity of a catalyst nanoparticle.
(b) Only the species A diffuses from the bulk, while the species B* is created at the nanoparticle surface. 
(c) Schematic representation of a yolk-shell nanoreactor. A nanoparticle of radius $R$ is embedded in a spherical shell of inner radius $\Rg$ and outer radius $\Rg+d$. 
The shell permeability depends on the diffusivity, $D_i(r)$, and on the transfer free energy profiles, $\DG_i(r)$. We model both as step functions with values $\bar{D}_i$ and $\DGbar_i$ inside, and $D_i^0$ and zero outside the shell, respectively.
}
\label{figYolkShell}
\end{center}
\end{figure*}
%%%%%%%%%%%%%%%%

We provide in this paper a theoretical description of {\it fully bimolecular}, surface-catalyzed reactions and, importantly, how they are connected and controlled to the individual reactant's permeabilities. We thus derive more comprehensive formulas of wider applicability for nanoreactors than the established unimolecular expressions. In our general scenario, {\it two} species A and B diffuse towards a catalyst nanoparticle, where they react together to produce another molecular species, C.
Our main result is the following formula for the total catalytic rate in bimolecular reactions:
\begin{align}
\ktot=
&\frac{1}{2}
\Bigg[
\frac{\kDA\kDB}{\kR}+\kDA+\kDB
\nonumber \\
&-\sqrt{\left(
\frac{\kDA\kDB}{\kR}+\kDA+\kDB
\right)^2-4\kDA\kDB}
\Bigg]
\eqs,
\end{align}
where $\kDA(\mathcal{P}\A)$ and $\kDB(\mathcal{P}\B)$ are the diffusion rates of the reactants A and B, which {\it explicitly} depend on the shell permeability, $\mathcal{P}_i$. As we will recall later, the latter is mathematically well defined within diffusion theory~\cite{Palasis:1992ew}.
This means that, in the general case, the diffusional mass fluxes of the different reactants are strongly {\it coupled}, providing a richer behavior than that arising in the simple unimolecular case. Only when one of the reactants has a much larger diffusion rate than the other, $\kDB\gg\kDA$, the total catalytic time reduces to the sum of the diffusion and the surface reaction times. If $\kDB\gg\kDA$ holds depends on permeability and thus on the nature and state of the nanoreactor shell.
We also study the case in which only one of the species, A, diffuses from the bulk, while the other one, B*, is created at the nanoparticle surface. We find that, with a proper redefinition of the diffusion rate, the total catalytic rate has the same form as in the case when both reactants diffuse from a bulk solution.
In the case of surface-controlled reactions, our theory converges to the low reactant adsorption limit of the Langmuir--Hinshelwood mechanism~\cite{Ertl:2008ie,Stamatakis:2012kx,Gu:2014cf}. With our theoretical description, we provide a firm and quantitative analysis of the conditions required for the limiting pseudo-unimolecular case to occur, and how they are controlled by nanoreactor permeability, clarifying when such simplified description is indeed valid. Based on our theory the permeability factor can be extracted from experiments or, if available from modeling or reference experiments, used to predict the outcome of nanoreactor reaction experiments.

In the remainder of the paper, we first introduce the theory in Section~\ref{sec:theory} and then proceed to discuss key results in Section~\ref{sec:results}, where we also present its application to the analysis of experimental data.

%%%%%%%%%%%%%%%%%%%%%%%%%%%%%%%%%%%%%%%%%%%%%%%%%%%%%%%%
\section{Theory for bimolecular reactions} \label{sec:theory}

We study surface-catalyzed bimolecular reactions in so-called yolk-shell (or `hollow-shell')  nanoreactors, depicted in Fig.~\ref{figYolkShell}. This hollow spherically symmetric configuration (see Fig.~\ref{figYolkShell}c) is the most general we can choose, since it covers catalytic nanoparticles confined in cavities as well as core-shell nanoreactors in the limit of zero gap ($\Rg=R$), i.e., no hollow compartment. We consider the two following scenarios:

\begin{itemize}
\item {\bf Case 1:}
Two species A and B diffuse from a bulk solution kept at constant concentration $c\AB^0$ towards the catalyst nanoparticle. A fraction (per unit of time and density of both reactants) of the reactants arriving at the surface, quantified by $\Kvol$, combines with each other to produce a third molecular species C (Fig.~\ref{figYolkShell}a).

\item
{\bf Case 2:} The species A and B combine at the nanoparticle surface as in the previous case, but only the species A diffuses from the bulk solution, whereas the species B (now denoted as B*) is generated locally at the surface (Fig.~\ref{figYolkShell}b).
\end{itemize}

Although it may appear artificial, the second case corresponds to the common situation where the nanoparticle surface catalyzes the production of reactive radicals of the species B* close to it, {e.g.} in photochemical reactions~\cite{Jia:2016cy}. These radicals rapidly decay away from the surface, so that molecules A can only react with B* in its proximity.

The total catalytic rate, $\ktot$, (number of molecules reacting per unit of time) is equal to the flux of reactants at the nanoparticle surface. In bimolecular reactions, the fraction of molecules A reacting is proportional to the number of molecules B at the same location, and vice versa. Thus, $\ktot$ can be also obtained as 
\begin{equation}\label{eq:ktotdef}
\ktot= \Kvol c\A(R) c\B(R)
\eqs,
\end{equation}
where $c\AB(R)$ are the reactants concentrations at the nanoparticle surface. Eq.~(\ref{eq:ktotdef}) is the low reactant adsorption limit of the Langmuir--Hinshelwood mechanism~\cite{Ertl:2008ie,Stamatakis:2012kx,Gu:2014cf}. To calculate $\ktot$, we solve the continuity equation for the density fields of reactants and product,
\begin{equation}\label{eq:cont}
\nabla \cdot \mathbf{J}_i =0
\eqs,
\end{equation}
$\mathbf{J}_i(r)$ being the flux of the species $i=$~A, B, C as a function of the distance from the nanoparticle.
In their diffusive approach to the catalyst nanoparticle (or when they diffuse away from it), reactants (products) have to permeate the shell. The kinetics of this process is thus governed by the shell permeability~\cite{Palasis:1992ew}, which depends on the diffusivity profile, $D_i(r)$, and on the thermodynamic barrier, i.e., the transfer free energy between bulk and shell, $\DG_i(r)$.
For simplicity, we take both profiles to be shell-centered step functions of width equal to the shell width $d$ (see Fig.~\ref{figYolkShell}c), {i.e.}
\begin{equation}\label{eq:difuprofile}
D_i(r)=
\begin{cases}
\bar{D}_{i} 	& \text{ \ } \Rg \leq r \leq \Rg+d  \eqs,\\
& \\
D_{i}^0 	& \text{ \ } \mathrm{elsewhere} \eqs,
\end{cases}
\end{equation}
and
\begin{equation}\label{eq:DGprofile}
\DG_i(r)=
\begin{cases}
\DGbar_i 	& \text{ \ } \Rg \leq r \leq \Rg+d  \eqs,\\
& \\
0 	& \text{ \ } \mathrm{elsewhere}\eqs.
\end{cases}
\end{equation}

Here, $\bar{D}_{i}$ and $D_{i}^0$ stand for the diffusion coefficient in the shell and solution, respectively. $\DGbar_i$ represents the mean interaction between the reactant and the shell (averaged over all molecular effects~\cite{Kanduc:2017bd}) and as such strongly depends on the state (swollen/collapsed) of the nanoreactor. 
Eqs.~(\ref{eq:difuprofile}) and (\ref{eq:DGprofile}) have been previously used as an approximation for spatially homogeneous gels~\cite{Lu:2011bi,CarregalRomero:2010gp,Herves:2012fp,Wu:2012bx,Jia:2016cy,AngiolettiUberti:2014hx}. 
Using standard thermodynamic relations~\cite{Dhont:1996ub}, we connect the flux of the species $i$ to its local concentration $c_i(r)$
\begin{equation}\label{eq:fluxes}
\mathbf{J}_i=-D_i \;\! c_i\nabla\beta\mu_i 
\eqs,
\end{equation}
where $\mu_i(r)$ is the chemical potential of the species $i$, and $\beta=1/\kB T$, with $\kB$ denoting the Boltzmann's constant and $T$ the absolute temperature of the system. 
The chemical potential of a molecule interacting with an external environment with a spatially dependent concentration and free energy is
\begin{equation}\label{eq:chempot}
%\beta\mu_i=\ln\left(\frac{c_i}{c_{i}^0}\right)+\beta\DG_i
%\eqs,
\beta\mu_i=\ln\left(\frac{c_i}{c_{i}^\mathrm{ref}}\right)+\beta\DG_i
\eqs,
\end{equation}
where $c_i^\mathrm{ref}$ is a reference concentration whose value can be chosen arbitrarily. For $\DG_i(r)=0$, Eq.~(\ref{eq:fluxes}) reduces to Fick's first law, $\mathbf{J}_i=-D_i \nabla c_i$. Using Eq.~(\ref{eq:chempot}), we can obtain $\DGbar_i$ by measuring the partitioning of the reactants $\mathcal{K}_i$, defined as the ratio between their concentrations inside and outside the shell, $c_i^\mathrm{in}$ and $c_i^0$, respectively:
\begin{equation}\label{eq:partitioning}
\mathcal{K}_i=\frac{c_i^\mathrm{in}}{c_{i}^0}=\mathrm{e}^{-\beta\DGbar_i}
\eqs.
\end{equation}

The partitioning in responsive polymer gels has been studied experimentally~\cite{Palasis:1992ew,Sassi:1996dj,Molina:2012iq} and in terms of computer modeling~\cite{QuesadaPerez:2014jk,Kanduc:2017bd} and theory~\cite{MonchoJorda:2014if}.
With the aforementioned definitions, the shell permeability is calculated as~\cite{Palasis:1992ew} 
\begin{equation}\label{eq:permeability}
\mathcal{P}_i=\bar{D}_i \mathcal{K}_i
\eqs.
\end{equation}

We calculate the steady-state distribution of reactants and products by solving Eq.~(\ref{eq:cont}) with the appropriate boundary conditions. 
The full mathematical derivation, including the boundary conditions we use for both cases studied, is included in the \SItext\  and here we just show the local reactants concentrations at the nanoparticle surface, $r=R$. In the case of a reactant diffusing from a bulk solution with a diffusion rate $k_{D\AB}$ we obtain
\begin{equation}\label{eq:reactantsconcR}
 c  \AB(R)
=
\frac{c  \AB^0 k_{D\AB}}
{\Kvol c  \AB^0 c  \BA(R) + k_{D\AB} \rme^{\beta \DG\AB(R)}}  
\eqs,
\end{equation}
and for a reactant B* created at the nanoparticle surface with a creation rate $K\BB^0$ we have
\begin{equation}\label{eq:reactantBconcR}
 c  \BB(R)
=
\frac{c\BB^0 K\BB^0}
{\Kvol c\BB^0  c\A(R) + K\BB^0 \rme^{\beta \DG\BB(R)}}
\eqs,
\end{equation}
with the permeability-dependent $c\BB^0$ defined as
\begin{equation}
c\BB^0=
\frac{K\BB^0}{4\pi}
\int^{\infty}_{R}
\frac{1}{\mathcal{P}\BB(r) \;\! r^2}\rmd r
\eqs.
\end{equation}

The total catalytic rate is calculated using Eq.~(\ref{eq:ktotdef}), combined with Eq.~(\ref{eq:reactantsconcR}) for the first case and with Eqs.~(\ref{eq:reactantsconcR}) and (\ref{eq:reactantBconcR}) for the second one. After some algebra (see \SItext) we obtain
\begin{align}\label{eq:totalrate}
\ktot=
&\frac{1}{2}
\Bigg[
\frac{\kDA\kDB}{\kR}+\kDA+\kDB
\nonumber \\
&-\sqrt{\left(
\frac{\kDA\kDB}{\kR}+\kDA+\kDB
\right)^2-4\kDA\kDB}
\Bigg]
\eqs
\end{align}
for both cases studied, {i.e.} the total catalytic rate has the same form regardless of the origin of the reactants (bulk or nanoparticle surface). In the previous expression,
\begin{equation}\label{eq:surfacerate}
\kR=\Kvol 
c\A^0 \rme^{-\beta \DG\A(R)}
c\B^0 \rme^{-\beta \DG\B(R)}
\eqs
\end{equation}
stands for the surface reaction rate, and
\begin{equation}\label{eq:difuratemono}
k_{D_i}=
4\pi c_i^0
\left[
\int^{\infty}_{R}
\frac{1}{\mathcal{P}_i(r) \;\! r^2}\rmd r
\right]^{-1}
\end{equation}
is the permeability-dependent diffusion rate of the reactant $i$ Case 1, whereas for Case 2 we need to redefine $\kDB=K\BB^0$.
%, $K\BB^0$  being the creation rate of reactants B* at the surface.
%
In the absence of shell, $\mathcal{P}_i(r) = D_i^0$, and the diffusion rate turns into the Smoluchowski rate~\cite{Smoluchowski:1917wh}  $k_{D_i}^0=4\pi R D_i^0c_i^0$.
For the yolk-shell configuration depicted in Fig.~\ref{figYolkShell}c the step profiles in Eqs.~(4) and (5) apply and the relation between the shell permeability and the diffusion rate, Eq.~(\ref{eq:difuratemono}), simplifies to 
\begin{equation}\label{eq:kDkD0}
\frac{k_{D_i}}{k_{D_i}^0}
=
\left[
1+\left(\frac{D_i^0}{\mathcal{P}_i}-1\right)
\left(\frac{R}{\Rg}-\frac{R}{\Rg+d}\right) 
\right]^{-1}
\!.
\end{equation}

%%%%%%%%%%%%%%%%%%%%%%%%%%%%%%%%%%%%%%%%%%%%%%%%%%%%%%%%
\section{Results and discussion} \label{sec:results}

Equation~(\ref{eq:totalrate}) is our main analytical result for surface-catalyzed bimolecular reactions. It shows that, in the fully bimolecular case, the diffusional fluxes of the different reactants are {\it coupled}. Thus, $\ktot$ depends in a non-trivial way on the surface and the diffusion rates and nanoreactor shell permeability, in contrast to the simple unimolecular case ({i.e.}, in general $\ktot^{-1}\neq k_D^{-1}+\kR^{-1}$). We also find that the total catalytic rate, once the diffusion rate is properly redefined, is described by Eq.~(\ref{eq:totalrate}) regardless of the origin of the reactants (bulk or nanoparticle surface).
For this reason, we will jointly analyze the results of both scenarios. In particular, we examine in detail the conditions required for the limiting pseudo-unimolecular case to occur. In addition, we use our theory to rationalize existing experimental data on the electron-transfer reaction between hexacyanoferrate (III) and borohydride ions at gold nanoparticles in responsive hydrogel-based nanoreactors~\cite{CarregalRomero:2010gp}.

\subsection{Pseudo-unimolecular reactions}

Bimolecular reactions are typically treated as pseudo-unimolecular when one of the reactants is in large excess with respect to the other~\cite{CarregalRomero:2010gp,Gu:2014cf,AngiolettiUberti:2015go,Wu:2012bx,Jia:2016cy}. 
The reasoning behind this assumption is that, according to the simple Smoluchowski rate, the reactant in larger concentration would diffuse towards the nanoparticle surface at a much larger rate than the other one. Therefore, when the reactant in limiting concentration arrives to the catalyst, it will always find a reactant of the other species to combine with.
However, this is not always true when considering nanoreactors. In this case, the diffusion rate, Eq.~(\ref{eq:difuratemono}), not only depends on the bulk reactant concentration but also on the shell permeability and thus on the molecular interactions of reactants with the shell. It is thus the combination of both quantities that determines whether a bimolecular reaction can be treated as pseudo-unimolecular, or not.

If one of the reactants has a much larger diffusion rate than the other one, {e.g.} $\kDB\gg\kDA$, the total reaction rate, Eq.~(\ref{eq:totalrate}), reduces to (see \SItext)
\begin{equation}\label{eq:totalrate-L2}
\ktot\rightarrow \ktot^1
=
\left({\kDA^{-1}}+\kR^{-1}\right)^{-1} 
\eqs,
\end{equation}
which is the well-known expression of the total reaction rate in unimolecular reactions, $\ktot^1$. In this case, the total catalytic time is the sum of the diffusion time of the slower reactant and the surface reaction time.
If both reactants diffuse from the bulk solution, according to Eq.~(\ref{eq:difuratemono}), this condition is satisfied when $c\B^0 \mathcal{P}\B \gg c\A^0 \mathcal{P}\A$. This means that one of the reactants should be in much higher bulk concentration and/or subjected to a much larger shell permeability than the other.
In the case where one of the reactants is created at the nanoparticle surface, we obtain $\kDB\gg\kDA$ if the creation rate of B* is much faster than the rate at which the reactants B* are transformed into products by reactants~A. 

%%%%%%%%%%%%%%%%FIG
\begin{figure}[t!]
\begin{center}
\includegraphics[width=\linewidth]{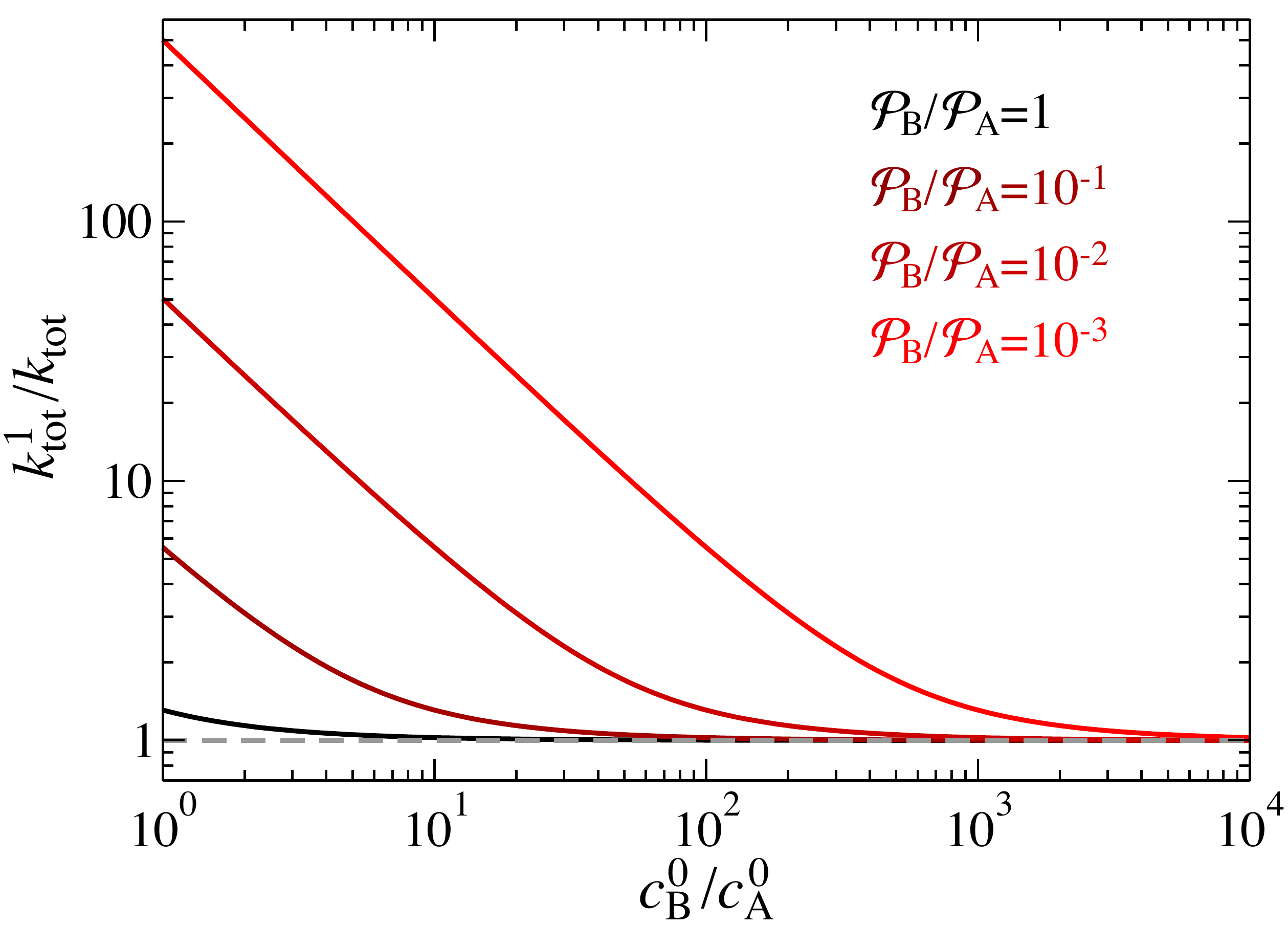}
\caption{Total rate for unimolecular reactions $\ktot^1$,  Eq.~(\ref{eq:totalrate-L2}), divided by the total reaction rate $\ktot$ for bimolecular reactions, Eq.~(\ref{eq:totalrate}), as a function of the relative reactant bulk concentration, $c\B^0/c\A^0$. The different lines stand for different relative shell permeabilities to the reactants, $\mathcal{P}\B/\mathcal{P}\A$. We assume $\kR=\kDA$ and a typical core-shell nanoreactor geometry with $d\gg R$.}
\label{figktot1ktot_conc}
\end{center}
\end{figure}
%%%%%%%%%%%%%%%%

In Fig.~\ref{figktot1ktot_conc} we analyze how large should be the excess of reactant B for the pseudo-unimolecular reaction limit to be valid. This value depends on the relative shell permeability, $\mathcal{P}\B/\mathcal{P}\A$.
For simplicity, we consider that the surface rate is equal to the diffusion rate of the reactant in limiting concentration ($\kR=\kDA$, diffusion-influenced reaction). We also consider a typical core-shell nanoreactor geometry with $d\gg R$. When both reactants have the same permeability (black line), the concentration of the reactant B should be roughly 10 times larger than the one of A to have a unimolecular reaction.
If we then decrease the shell permeability to the reactant B 10 times, its concentration has thus to become 100 times higher with respect to that of A to keep this limit (moving from darker to lighter red lines).
Fig.~\ref{figktot1ktot_conc} also shows that the catalytic rate predicted for a pseudo-unimolecular reaction for the reactant in limiting concentration may differ from the fully bimolecular one by orders of magnitude. Thus, when dealing with nanoreactors, it is necessary to consider not only the difference between the bulk reactants concentration but also the difference in the shell permeability to the reactants.

%%%%%%%%%%%%%%%%FIG
\begin{figure}[t!]
\begin{center}
\includegraphics[width=\linewidth]{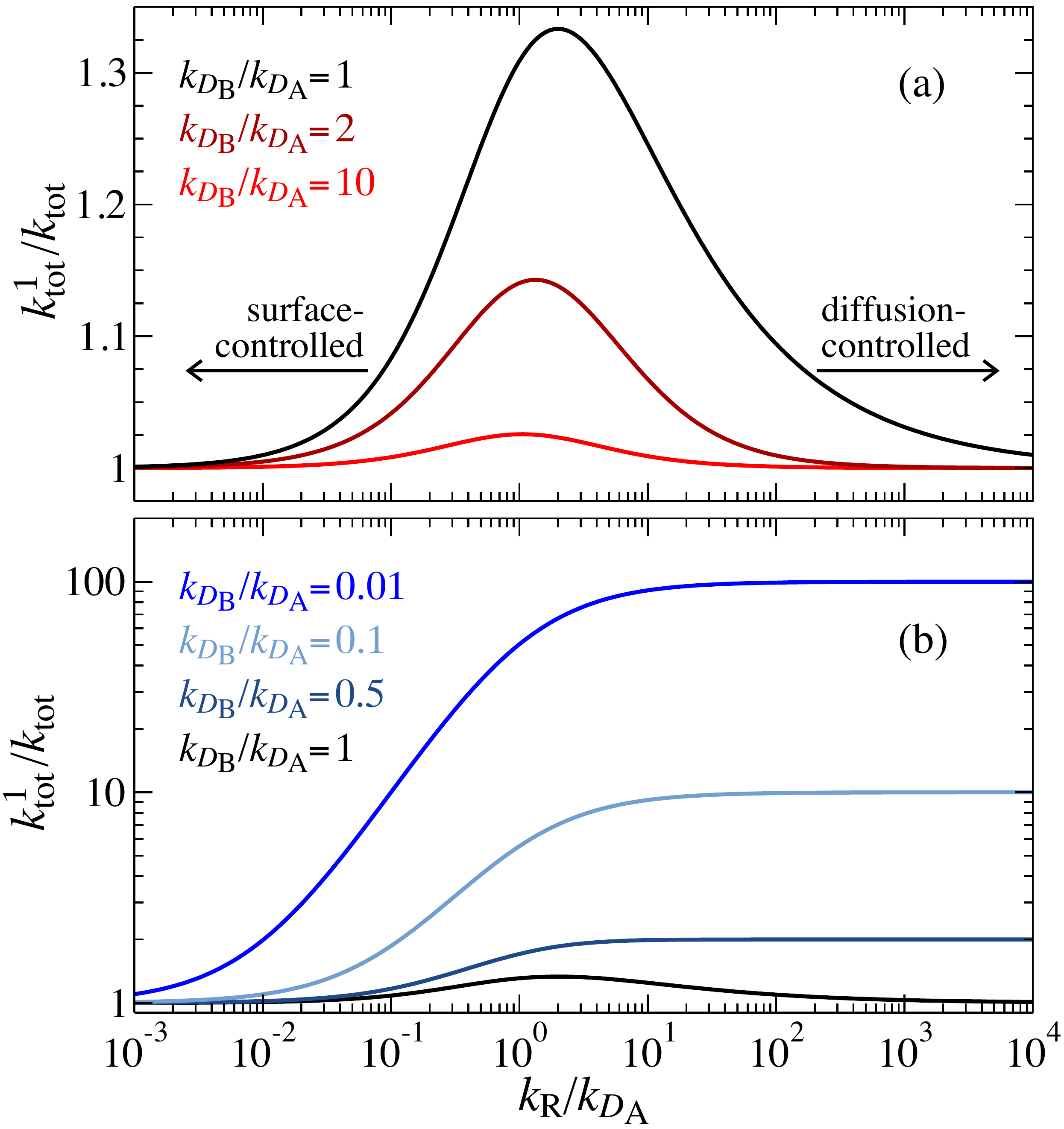}
\caption{Total rate for unimolecular reactions $\ktot^1$,  Eq.~(\ref{eq:totalrate-L2}), divided by the total reaction rate $\ktot$ for bimolecular reactions, Eq.~(\ref{eq:totalrate}), as a function of the reduced surface rate values, $\kR/\kDA$. The different lines stand for different relative reactant diffusion rates $\kDB/\kDA$. The reddish lines (a) consider that the reactant A diffuses at a slower rate than B. The bluish lines (b) consider the opposite case. The case of equal reactant diffusion rate is shown by the black lines in (a) and (b).}
\label{figktot1ktot_difu}
\end{center}
\end{figure}
%%%%%%%%%%%%%%%%

As a consequence, in Fig.~\ref{figktot1ktot_difu}a we study how large should be the diffusion rate of the reactant B in comparison with the one of the reactant A ({i.e.}, how large should be the ratio $\kDB/\kDA$) in order for the pseudo-unimolecular reaction limit to be valid. This critical value depends on the ratio between the surface rate and the diffusion rate of the slower reactant, $\kR/\kDA$.
If $\kR\sim\kDA$ (diffusion-influenced reaction), 
%we conclude that $\kDB\gg \kDA$ to simplify the problem to the one of a pseudo-unimolecular reaction. 
we observe that $\kDB$ must be about an order of magnitude larger than $\kDA$ to simplify the problem to the one of a pseudo-unimolecular reaction within a $2\%$ of accuracy. 
If both reactants diffuse at the same rate, the total reaction rate estimated by Eq.~(\ref{eq:totalrate-L2}) is $\sim$ 30$\%$ larger than the real one.

In the previous discussion we assumed that A is the limiting reactant and, consequently, we calculated the catalytic rate for unimolecular reactions given by Eq.~(\ref{eq:totalrate-L2}) using its diffusion rate. In Fig.~\ref{figktot1ktot_difu}b (the inset) we show that if the choice of the limiting reactant is wrongly made  
(i.e., B is the limiting one), the predicted unimolecular rate can be more than one order of magnitude larger than the real one. The plateau reached at larger surface rates coincides with the value of the unimolecular rate considering B as the limiting reactant.
We would like to emphasize here that in stimuli-responsive nanoreactors the physico-chemical properties of the shell can drastically change around a critical value of the external stimulus, {e.g.} from hydrophilic to hydrophobic in PNIPAM-based nanoreactors at the critical temperature. These changes strongly affect the permeability, and can thus switch the identity of the limiting reactant from one type to another. Figure 3b thus shows that by not recognizing this fact when analyzing experiments, theoretical predictions of the rates can be off by order of magnitude.

Figure~\ref{figktot1ktot_difu}a also shows that, if the limiting reactant is properly identified, we can always treat a bimolecular reaction as unimolecular if the reaction is diffusion- or surface-controlled. 
This can be also inferred from Eq.~(\ref{eq:totalrate}). 
On the one hand, if the diffusion rate of the slowest reactant is much larger than the surface rate, $\kDA \gg \kR$, the total catalytic rate turns into $\ktot \rightarrow \kR$, which agrees with the surface-controlled limit for a unimolecular reaction, Eq.~(\ref{eq:totalrate-L2}).
On the other hand, if we take the limit of very fast surface rate, $\kR\rightarrow \infty$, we obtain (see \SItext)
\begin{equation}\label{eq:totalrate-L1}
\ktot \rightarrow
\begin{cases}
\kDA		& \text{\ if\ \ } \kDA<\kDB  \eqs, \\
& \\
\kDB 	& \text{\ if\ \ }  \kDA>\kDB  \eqs.
\end{cases}
\end{equation}

This means that if the reaction at the surface of the catalytic nanoparticle is immediate, then what matters is only the diffusion time of the slower reactant. 
Thus, the reaction becomes pseudo-unimolecular and diffusion controlled.
From a physical point of view,  the explanation is as follows: There are so many `active' reactants at the nanoparticle surface that one of them will necessarily react whenever a reactant of the other species arrives.

\subsection{Application to bimolecular reactions in responsive core-shell nanoreactors}\label{subsec:exp}

%%%%%%%%%%%%%%%%FIG
\begin{figure}[t!]
\begin{center}
\includegraphics[width=\linewidth]{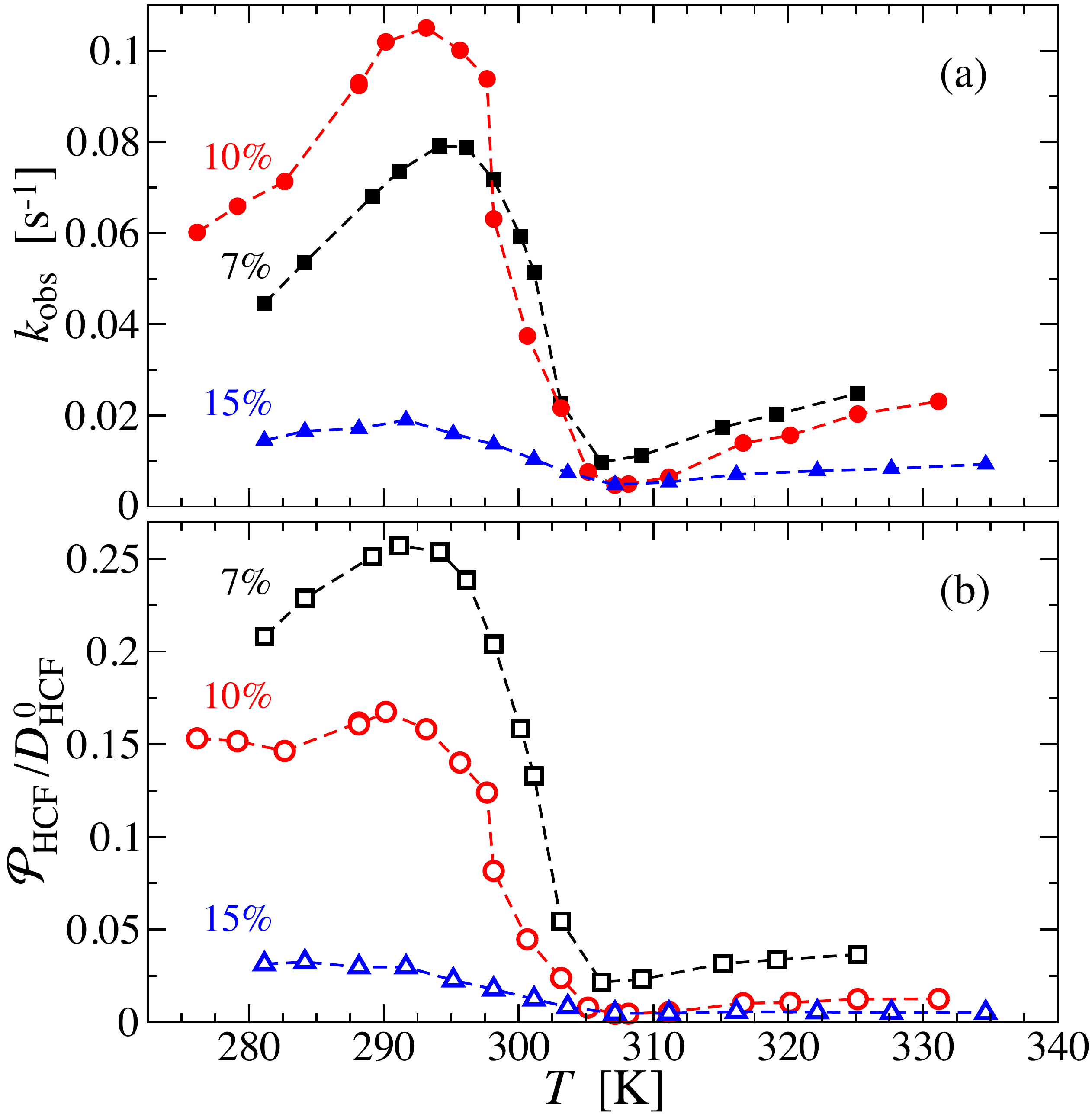}
\caption{
(a) Influence of the temperature on the measured pseudo-first order constant in the Au-PNIPAM nanoreactors for different cross-linking densities: ($\blacksquare$) $7\%$ BIS,  ($\bullet$)$10\%$ BIS, and ($\blacktriangle$) $15\%$ BIS. Catalytic data taken from Ref.~\cite{CarregalRomero:2010gp}.
(b) Influence of the temperature on the permeability of the PNIPAM shell to HCF, $\mathcal{P}_\mathrm{HCF}/D^0_\mathrm{HCF}$, estimated using Eq. (\ref{eq:permD0}), for different cross-linking densities: ($\Box$) $7\%$ BIS, ($\circ$)$10\%$ BIS, and $(\triangle$) $15\%$ BIS. 
%The solid colored lines for the swollen nanoreactor state are obtained using Eq.~(\ref{eq:permeability}) with the values from Table~\ref{tbl:datafit}.
}
\label{figPermT}
\end{center}
\end{figure}
%%%%%%%%%%%%%%%%

Carregal-Romero {et al.} investigated the bimolecular electron-transfer reaction between hexacyanoferrate (III), $\mathrm{Fe(CN)}_6^{3-}$ (HCF), and borohydride, $\mathrm{BH}_4^-$, ions in Au-PNIPAM core-shell nanoreactors~\cite{CarregalRomero:2010gp}.
In Fig.~\ref{figPermT}a we show the experimentally measured temperature-dependent catalytic rate for different N,N'-Methylenebisacrylamide (BIS) cross-linking densities. We first demonstrate that the bimolecular reaction can be treated as pseudo-unimolecular and is diffusion-controlled. Then we show how our theoretical framework can be used to analyze the experimental data and dissect the permeability into diffusion and partitioning effects, thereby providing an improved understanding of all underlying mechanistic effects in this system. 

As we pointed out before, it is the combination of bulk reactant concentration and permeability what determines if a bimolecular reaction can be treated as pseudo-unimolecular.
The ratio between the bulk borohydride and HCF concentrations is $c^0_{\mathrm{BH}_4^-}/c^0_\mathrm{HCF}=125$ ($c^0_{\mathrm{BH}_4^-}=50$ mM, $c^0_\mathrm{HCF}=0.4$ mM). Therefore, according to Fig.~\ref{figktot1ktot_conc}, the permeability ratio should be, $\mathcal{P}_{\mathrm{BH}_4^-}/\mathcal{P}_\mathrm{HCF}\gtrsim 0.1$, for the pseudo-unimolecular reaction limit to be valid. 
This is easily fulfilled as HCF, in simple terms, is bigger and slower than the borohydride. Partitioning of HCF has been measured in microgels (with lower crosslink density of 5\%)\cite{Mergel:2014cu} and indeed found to be weakly excluded, i.e. clearly $\mathcal{K}_\mathrm{HCF} < 1$, while for the simple and small ion borohydride excluded volume effects must be expected to be smaller. In addition, the larger HCF diffuses slower than borohydride due to more steric `obstruction' effects.\cite{Masaro:1999ep} Thus, HCF to a high certainty has a smaller shell permeability than borohydride, which means that the reaction can be treated as pseudo-unimolecular, with a total rate given by Eq.~(\ref{eq:totalrate-L2}), being HCF the limiting reactant.
In this limit, the relation between the experimentally observed rate and the calculated total rate is $k_\mathrm{obs}=\ktot({c^0_\mathrm{NR}}/{c^0_\mathrm{HCF}})$~\cite{AngiolettiUberti:2015go}, where $c^0_\mathrm{NR}$ stands for the nanoreactor concentration.
 
To discern if the reaction is diffusion- or surface-controlled, we estimate the observed Smoluchowski rate, $k_{D_\mathrm{HCF}}^0({c^0_\mathrm{NR}}/{c^0_\mathrm{HCF}})\approx 0.3-0.7\ \mathrm{s}^{-1}$ (see \SItext\ or Ref.~\cite{CarregalRomero:2010gp} for the input experimental values). According to Fig.~\ref{figPermT}a, $k_\mathrm{obs}\approx 0.2 \ \mathrm{s}^{-1}$ in the most `open state of the shell, i.e., being swollen at the smallest cross-linker density.  This means the reaction is diffusion controlled (consistent with common literature assumptions~\cite{CarregalRomero:2010gp,Herves:2012fp,Freund:1959bj}), i.e. $k_\mathrm{obs}=k_{D_\mathrm{HCF}}({c^0_\mathrm{NR}}/{c^0_\mathrm{HCF}})$, with $k_{D_\mathrm{HCF}}$, given by Eq.~(\ref{eq:difuratemono}),  and the smaller observed rates for larger crosslinker-densities and collapsed states must be ascribed to the decreasing shell permeability to HCF. 

Now, using Eq.~(\ref{eq:kDkD0}), we obtain the shell permeability to HCF, $\mathcal{P}_\mathrm{HCF}$, as
\begin{equation}\label{eq:permD0}
\mathcal{P}_\mathrm{HCF}
=
D^0_\mathrm{HCF} \left[
1+\left(\frac{k_D^0}{k_D}-1\right)\left(1-\frac{R}{R+d}\right)^{-1}
\right]^{-1}
\!,
\end{equation}
where we have already considered that the nanoreactor has a core-shell geometry, $\Rg=R$. 
%The relation between the diffusion rate and the experimentally observed rate is $k_D=k_\mathrm{obs}({c^0_\mathrm{HCF}}/{c^0_\mathrm{NR}})$~\cite{AngiolettiUberti:2015go}, where $c^0_\mathrm{NR}$ stands for the nanoreactor concentration.
$D^0_\mathrm{HCF}=\kB T/6\pi\eta(T) a$ represents the bulk diffusion coefficient of HCF with $\eta(T)$ being the temperature-dependent viscosity of water and $a\approx 0.3$~nm the HCF size~\cite{CarregalRomero:2010gp}. That is, the $T$-dependence of the bulk diffusion is explicitly considered in our analysis. 

We plot the temperature-dependent shell permeability to HCF calculated using Eq.~(\ref{eq:permD0}) for different cross-linking densities in Fig.~\ref{figPermT}b. We make use of the catalytic data ($k_\mathrm{obs}$, shown in Fig.~\ref{figPermT}a) and hydrodynamic radius data ($r_\mathrm{H}\approx R+d$, shown in Fig.~S1b in the \SItext) measured by Carregal-Romero {et al.}~\cite{CarregalRomero:2010gp}.  
We see already that the measured catalytic rate, $k_\mathrm{obs}$, does not follow the same trend with the cross-linking density, Fig.~\ref{figPermT}a, than the permeability, Fig.~\ref{figPermT}b, because each system was prepared at a different nanoreactor concentration. The trend is recovered when the observed rate is normalized by $c^0_\mathrm{NR}$ (Fig.~S1a in the \SItext).
Fig.~\ref{figPermT}b also shows that the shell permeability to HCF switches relatively sharply at the polymer volume transition and is about one order of magnitude larger in the swollen than in the collapsed state of the nanoreactor, which gives rise to a larger catalytic rate, Fig.~\ref{figPermT}a. 

We further find that all the permeability values are below the HCF reference permeability in bulk solvent, i.e. $\mathcal{P}_\mathrm{HCF}/D^0_\mathrm{HCF}<1$  (with $D^0_\mathrm{HCF}\approx 0.9$ nm/ns at 298.2 K~\cite{CarregalRomero:2010gp}), which, according to Eq.~(\ref{eq:permeability}), must be assigned to small partitioning and/or slow diffusion.  We examine this in the following in even more detail,  while we solely focus on the swollen part as only for this case further meaningful quantitative assumptions on diffusion can be made.  In the swollen state, the  mobility decrease can be described by obstruction-hindered diffusion, e.g., as expressed by  
\begin{equation}\label{eq:difuMM}
\frac{\bar{D}_i}{D^0_i}
=
 \left(
\frac{1-\phi}{1+\phi}
\right)^2
\!,
\end{equation}
being $\phi$ the PNIPAM volume fraction in the shell. This expression was developed by Mackie and Meares~\cite{Mackie:1955if} to explain diffusion due to excluded-volume effects of the polymer and it is valid for small-sized solutes in semi-dilute polymer solutions~\cite{Masaro:1999ep}, i.e. it should be applicable to the swollen state of the hydrogel. 
%
%The second effect must be assigned to a positive transfer free energy, $\Delta \bar{G}_\mathrm{HCF}>0$,  i.e.,  following Eq.~(\ref{eq:partitioning}) a depleted HCF partitioning $\mathcal{K}_\mathrm{HCF}<1$. A depletion 
%$\mathcal{K}_\mathrm{HCF}\simeq 0.5$ was recently measured in cationic hydrogels~\cite{Mergel:2014cu} with similar cross-linking densities as the systems under consideration here. HCF is a trivalent anion and a depletion 
%to a cationic hydrogel points to a dominant repulsion from the hydrogel due to effects other than simple charge attraction, e.g., excluded volume and image charge repulsion (as the shell in general has a lower dielectric constant than bulk water).  The charge state of the system studied by Carregal-Romero {et al.} was not reported and was probably being near electroneutral. Hence, a similar or even larger (due to missing electrostatic attraction) depletion in the hydrogel can be expected here. These diffusion and partitioning effects also explain the decrease in permeability with increasing cross-linking density, for which the excluded-volume effect grows and the mean dielectric constant of the shell decreases. 

We now separate out the diffusivity from the permeabilities to extract more quantitative numbers of the partitioning for further discussion. For the swollen state we assume the diffusivity is fully provided by Eq.~(\ref{eq:difuMM}) including its $T$-dependence.  For this, we calculate the PNIPAM volume fraction for different cross-linking densities as
\begin{equation}\label{eq:volfrac}
\phi=\frac{V_\mathrm{P}}{V}=\frac{m_\mathrm{P}/\rho^0_\mathrm{P}}{V}= \frac{\rho_\mathrm{P}}{\rho^0_\mathrm{P}}
\eqs,
\end{equation}
where $V$ stands for the whole suspension volume, $V_\mathrm{P}$ is the volume occupied by the PNIPAM in the shell, $m_\mathrm{P}$ its mass, $\rho^0_\mathrm{P}=1.1$ g/cm$^3$~\cite{Haynes:2011ue} the mass density of a PNIPAM polymer, and $\rho_\mathrm{P}$ the mean PNIPAM segment density at different cross-linking densities~\cite{Varga:2001bm}. 
The fact that the normalized permeabilities, $\mathcal{P}_\mathrm{HCF}/D^0_\mathrm{HCF}$,  are approximately constant at low temperatures indicates that the increment of the permeability with the temperature in the swollen state of the nanoreactor is mostly due to the diminution of the water viscosity with the temperature. This is not the case for the 7\% BIS case since a fully swollen nanoreactor was not achieved in the experimental conditions (see Fig.~S1b in the \SItext).

\begin{table}[t!]
\small
  \caption{
  Mean PNIPAM segment density, $\rho_\mathrm{P}$, obtained from Ref.~\cite{Varga:2001bm} (the values at $10\%$ and $15\%$ BIS are calculated using a linear extrapolation); PNIPAM volume fraction in the shell, $\phi$, using Eq.~(\ref{eq:volfrac}); ratio between the HCF diffusivities inside and outside the shell, ${\bar{D}_\mathrm{HCF}}/{D^0_\mathrm{HCF}}$, according to Eq.~(\ref{eq:difuMM}); HCF partitioning, $\mathcal{K}_\mathrm{HCF}$, to fit the permeability data from Fig.~\ref{figPermT}b in the swollen nanoreactor state using Eq.~(\ref{eq:permeability}); and HCF transfer free energy, $\beta\Delta \bar{G}_\mathrm{HCF}$, obtained by Eq.~(\ref{eq:partitioning}). Values shown for different cross-linking densities in the swollen nanoreactor state.}
  \label{tbl:datafit}
  \renewcommand{\arraystretch}{1.5}
  \begin{tabular*}{0.48\textwidth}{@{\extracolsep{\fill}}cccccc}
    \hline 
    $\%$ BIS & $\rho_\mathrm{P}$ [g/cm$^3$] & $\phi$ & $\frac{\bar{D}_\mathrm{HCF}}{D^0_\mathrm{HCF}}$ & $\mathcal{K}_\mathrm{HCF}$ & $\beta\Delta \bar{G}_\mathrm{HCF}$ \\ 
    \hline
    $7$ 	  & 0.14 & $0.13$ 	& $0.6$ 	& $0.4$ 	& $+1.0$	\\
    $10$ 	  & 0.19 & $0.17$ 	& $0.5$	& $0.3$ 	& $+1.2$	\\
    $15$ 	  & 0.27 & $0.24$		& $0.4$	& $0.08$	& $+2.5$	\\
    \hline
  \end{tabular*}
\end{table}

Using Eq.~({\ref{eq:permeability}}), we calculate the necessary partitionings to fit the permeabilities in the swollen nanoreactor state. The results are given in Table~\ref{tbl:datafit}.
%
%The solid colored lines in Fig.~\ref{figPermT}a are fittings to the permeability for different cross-linking densities in the swollen state using Eq.~(\ref{eq:permeability}) with the values from Table~\ref{tbl:datafit}. 
%
%{\color{blue}We do not include a fit for the 7\% BIS case since it is not clear how to define the swollen nanoreactor case for such cross-linking density (see Fig. S1b in the \SItext).}
%
%The fact that the {\it slopes} of the fitting lines agree well with the permeability data indicates that the increment of the permeability with the temperature in the swollen state of the nanoreactor is mostly due to the diminution of the water viscosity with the temperature.
%
Importantly, the HCF transfer free energies, predicted using  Eq.~(\ref{eq:partitioning}), are all positive and increase with the cross-linking density. This trend is expected as the excluded-volume of the shell increases with less water content and less free space. Values of the partitioning are reasonably close to the ones recently measured for HCF for cationic hydrogels.\cite{Mergel:2014cu}
Hence, our theory allows a full dissection of experimental rate and permeability effects into basic underlying mechanisms, such as bulk diffusion,  hindered diffusion in the shell and thermodynamic partitioning effects based on physical interactions. Clearly, more work is in order to understand details of molecular interactions and mobility, in particular in the collapsed state of the hydrogels.

%%%%%%%%%%%%%%%%%%%%%%%%%%%%%%%%%%%%%%%%%%%%%%%%%%%%%%%%
\section{Conclusions}

In this paper, we presented the rate theory for surface-catalyzed {\it bimolecular} reactions in responsive nanoreactors. 
Our theory for nanoreactors is markedly different from the treatment of standard bimolecular reactions in bulk, since the reactions here only occur at the surface of the catalyst, coupling the surface reaction rate with the permeability of the shell, which can be controlled and designed by material synthesis and stimuli in the environment. Our theory allows to extract permeabilities from experiments, or, alternatively, if the permeabilities are available from reference experiments or modeling, can be employed to predict and thus rationally design nanoreactors kinetics. 

We found that the total rate for bimolecular reactions depends in a non-trivial way on the reactants diffusion and surface reaction rates. More precisely, the coupling between these timescales makes it impossible to separate the process into distinct `diffusion+reaction' steps, as is the case for standard pseudo-unimolecular descriptions previously used to rationalize experiments (interestingly, breakdown of the standard theory can also be observed in the case of strong coupling with shell fluctuations~\cite{Kolb:2016jg}).
 
In this regard, most of the studies in the literature assume to be dealing with pseudo-unimolecular reactions, usually by working with one of the reactants in large excess. We show that this might not be enough: Tighter conditions strongly dependent on shell permeability are required, whose breakage can lead to large error in the analysis of kinetics. This indicates that in nanoreactors one can in principle switch between pseudo-unimolecular and bimolecular reactions, or change the identity of the limiting reactant, by temperature and other stimuli. In addition, we also found that, with a proper redefinition of the diffusion rate, the total catalytic rate assumes the same form regardless of the origin of the reactants (bulk or nanoparticle surface).
  
As a practical demonstration we applied our theory to available experimental data on the bimolecular electron-transfer reaction between HCF and borohydride ions in Au-PNIPAM core-shell nanoreactors~\cite{CarregalRomero:2010gp}. A thorough analysis of the permeability showed that the rate is governed by a complex interplay between diffusion and partitioning effects that define the shell permeability to HCF. 
  
We finally note that our theory for bimolecular reactions is strictly valid for low adsorption of reactants only. Thus, a theory accounting for diffusion-influenced Langmuir--Hinshelwood mechanisms~\cite{Miura:2015kp} in responsive nanoreactors is still needed. All these features will highlight even richer behavior and dynamic control achievable in responsive nanoreactors. 
\newline

\emph{Supporting Information Available}: Plot of the temperature-dependent observed reaction rate per nanoreactor concentration, $k_\mathrm{obs}/c^0_\mathrm{NR}$, and hydrodynamic radius for different cross-linking densities (data taken from Ref.~\cite{CarregalRomero:2010gp});  derivation of the expressions for the total catalytic rate and for the reactants and products distributions. %This material is available free of charge on the ACS Publications website at http://pubs.acs.org.

\begin{acknowledgments}
The authors thank Matthias Ballauff, Yan Lu, and Daniel Besold for helpful discussions on catalytic nanoreactors experiments. 
This project has received funding from the European Research Council (ERC) under the European Union's Horizon 2020 research and innovation programme (grant agreement n$^\circ$ 646659-NANOREACTOR).
%R.R., W.K.K., M.K. and J.D. acknowledge funding by the ERC (European Research Council) Consolidator Grant with project number 646659-NANOREACTOR. 
%
S.A.-U. acknowledges financial support from the Beijing Advanced Innovation Centre for Soft Matter Science and Engineering.
%
%We thank the North-German Supercomputing Alliance (HLRN) for providing computing resources.

\end{acknowledgments}

%%%%BIBLIOGRAPHY
\bibliography{bimolreact_refs}
\bibliographystyle{apsrev4-1}

%\bibliographystyle{apsrev4-1}
%\bibliography{filtration_refs.bib}

\end{document}